# Giant exchange bias and ferromagnetism in the CoO shell of Co/CoO-MgO core-shell nanoparticles

C. N. Ge[1,2], Xiangang Wan[1,*], E. Pellegrin[3,**], Z. Hu[4], W. Q. Zou[1,***], Y. W. Du[1]

[1] *National Laboratory of Solid State Microstructures and Department of Physics, Nanjing University, Nanjing 210093, China*

[2] *Department of Physics, Jiangsu Institute of Education, Nanjing 210013, China*

[3] *CELLS-ALBA Experiments Division, Carretera BP 1413, km 3.3, E-08290 Cerdanyola del Vallès (Barcelona), Spain*

[4] *Max Planck Institute for Chemical Physics of Solids, Nöthnitzer Straße 40, D-01187 Dresden Germany*

## ABSTRACT

Using magnetron sputtering, we produced a series of Co/CoO-MgO nanoparticles on Si(100) substrates. High-resolution transmission electron microscopy (HRTEM) image shows that small isolated Co-clusters (core) covered with CoO (shells) with a size of a few nm embedded in a MgO matrix. Resistivity as a function of Co atomic ratio exhibits a distinct percolation threshold with a sharp decrease around 69% Co content. Across the threshold, the resistivity drops about 7 orders of magnitude. For a sample at this percolation critical threshold, we have observed a giant exchange bias field $H_E \approx 2460$ Oe at T= 2K, and using soft x-ray magnetic circular dichroism at the Co-$L_{2,3}$ edge, we have detected a ferromagnetic (FM) signal originating from the antiferromagnetic CoO shell. Moreover, decreasing the Mg-impurities will reduce the FM signal from CoO shell (namely the uncompensated spin density) and the size of $H_E$, thus directly support the uncompensated spin model.

PCAS numbers: 75.75.-c, 75.70.cn

Exchange bias (EB) was first reported by Meiklejohn and Bean in the system of partially oxidized Co particles [1]. The outstanding characteristic of EB is a hysteresis loop shifting with the magnetic field, which is commonly accompanied with an increase of the coercivity ($H_C$) and the appearance of an unidirectional anisotropy. EB has already been widely used in magnetoresistive reading heads [2,3], as well as spin-valve-based devices [4], and has also been proposed as an efficient way to stabilize the written information against thermal fluctuations in magnetic recording media [5]. Although tremendous efforts have been devoted to understand this intriguing phenomenon and to find out new systems for various applications [6-14], the microscopic origin of the EB effect is still controversially discussed in terms of uncompensated(UC) interfacial spins [13,14,15], spin flop coupling within the ferromagnetic (FM) and antiferromagnetic (AFM) layers [16,17] and spin canting within the AFM system induced by the exchange coupling between FM and AFM [5].

---

[*] Electronic mail:  xgwan@nju.edu.cn
[**] Electronic mail:  epellegrin@cells.es
[***] Electronic mail:  wqzou@nju.edu.cn

It was found that Co/CoO core-shell system has a larger exchange bias field ($H_E$) than the Co/CoO bilayer system [5]. This could obviously be interpreted as an increase of the amount of interface surface and interface roughness between the Co core and the CoO shell, supporting both the interfacial uncompensated spin model as well as the spin canting within the AFM CoO shell. For the Co/CoO core-shell or Co/CoO bilayer systems, the EB is primarily due to the roughness of the interface [18,19]. The associated uncompensated interfacial spins and the number of canted spins are not easy to measure, thus a precise experimental analysis of the microscopic origin of EB is still lacking. Previously, it was found that in Co/CoO bilayered thin films $H_E$ can be significantly increased if the AFM CoO layer is diluted with Mg ions [20,21]. On the other hand, Skumryev *et al.* [5] have found that Co nanoparticles embedded in a CoO matrix have a larger $H_E$ and $H_C$ than when embedded in other matrix. Therefore for the core-shell system, if the Co atoms in AFM CoO are diluted by Mg substitution, we can expect an increase of the presumably uncompensated spin within the AFM CoO shell. Thus, the enhancement of uncompensated spin density in the CoO shell provides an opportunity to observe the sizeable FM signals from the CoO shell using soft x-ray magnetic circular dichroism at the Co-$L_{2,3}$ edges. The experimental observation of the existence of the latter will provide a direct evidence of the uncompensated spins. Until now and according to our present knowledge, there are no such experimental data available.

In this work, a series of Co/CoO-MgO (CCMO) granular films were deposited on Si(100) substrates by magnetron sputtering under a vacuum pressure of $2\times10^{-7}$ mbar. The sputtering targets were mosaic-like, made by adhering small square pieces of highly purified cobalt metal onto a MgO plate. All the Co/CoO-MgO granular films with different metal atomic ratio were prepared at room temperature using those sputtering targets. In the magnetically controlled sputtering process, Co atoms were polymerized and formed into metallic clusters. During the same process, the outside of the Co clusters was oxidized, thus forming the CoO shell. Considering that MgO and CoO share the same NaCl crystallographic structure and their lattice parameters differ by only 1.1% [22], some Co ions inside the CoO shell were replaced by Mg. The exact composition of all samples was investigated by energy dispersive x-ray spectroscopy (EDX). The resistance of all samples was measured using the four-terminal measuring technique. The magnetic characterization of the samples was performed by a Quantum Design superconducting quantum interference device (SQUID) magnetometer.

The x-ray magnetic circular dichroism (XMCD) spectra at the Co-$L_{2,3}$ edges were collected at the BL29 Boreas beamline at the CELLS-ALBA synchrotron radiation facility in Barcelona with a resolution of 0.25 eV and a degree of circular polarization close to 100% in a magnetic field of 5 Tesla and a sample temperature of 80K. The spectra were recorded using the total electron yield method (by measuring the sample drain current) in a chamber with a vacuum base pressure of $2\times10^{-10}$ mbar.

Firstly, we show the room temperature resistance data as a function of the Co atomic ratio in Fig. 1. The resistance drops by about 7 orders of magnitude from 55% to 80% Co atomic ratio. There is a distinct percolation threshold around 69% Co atomic ratio (i.e., $Co_{69}Mg_7O_{24}$ called CCMO1 below) as clearly shown in Fig.1. We can explain the sharp drop in resistance as follows: when the Co atomic ratio is less than 69%, the Co/CoO particles are well separated and the electron hopping between them is small, consequently the system has a high resistance and exhibits insulating behavior. When increasing the Co content, the hopping between core/shell particles becomes possible and the resistance decreases. In case the Co concentration is larger than the percolation threshold value, hopping between the core/shell particles is significantly increased and the Co/CoO-MgO system exhibits metallic properties. The CCMO1 sample with 69% Co atomic ratio is located at the critical point and is thus expected to have some peculiar properties. Therefore, we will concentrate on this sample from now on.

The morphological characterization of the samples was revealed by HRTEM. Figure 2 shows the typical image of the CCMO1 sample, where small Co clusters covered with a CoO shell embedded in a MgO matrix can be observed. The mean diameter of Co clusters is about 4-5 nm and the thickness of the CoO shell is about 1-2 nm. Notice that these Co/CoO particles are isolated by a nonmagnetic MgO matrix. Obviously, the sample is polycrystalline, and the small Co/CoO particles are island-like. The shape of those Co atomic clusters is irregular as shown in Fig. 2, which means that the interface between the FM Co cluster core and the AFM CoO shell is rough.

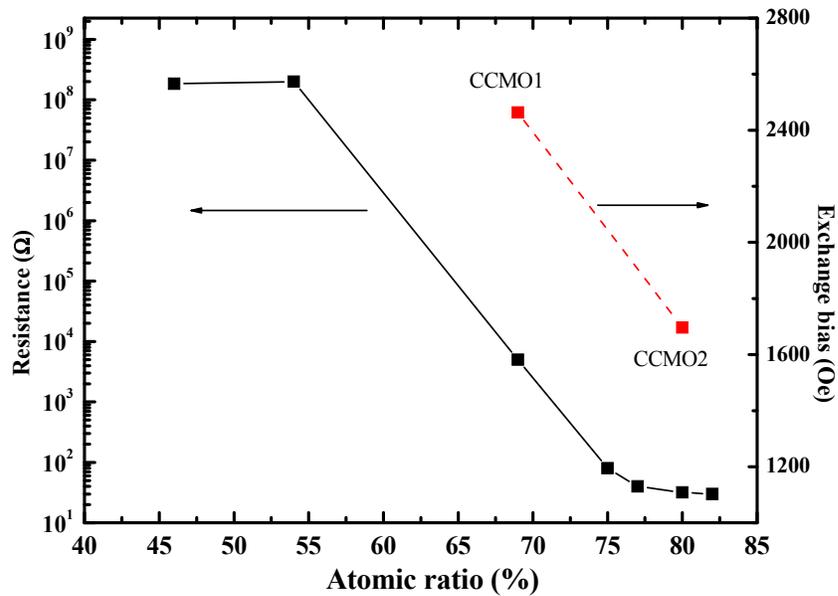

Fig. 1. Resistance (black squares) as a function of the Co atomic ratio at room

temperature. The red squares denote the exchange bias field $H_E$ of CCMO1 and CCMO2 at 2K.

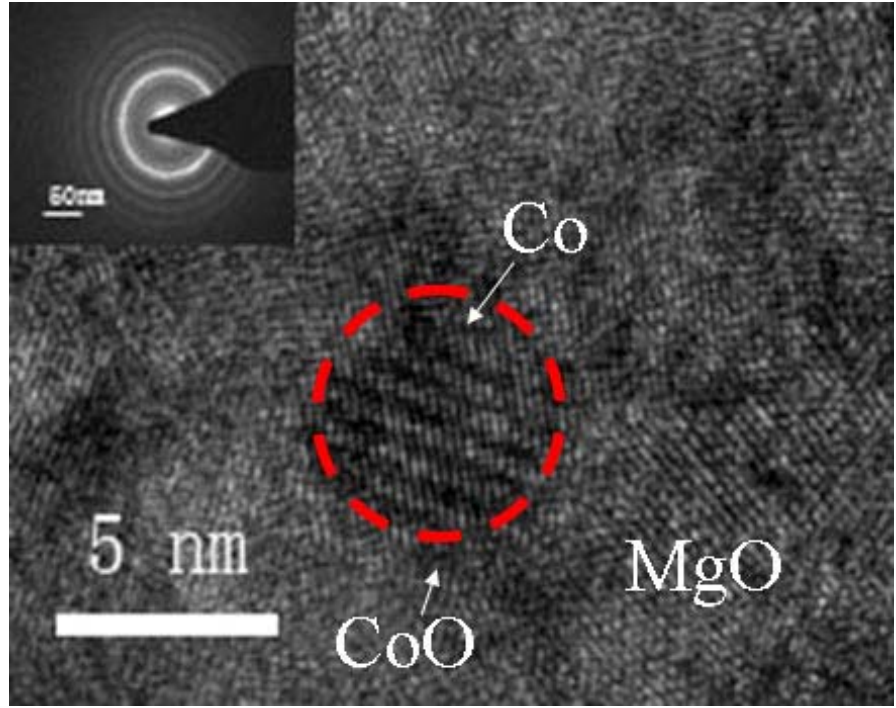

Fig. 2. Transmission electron micrographs (TEM) showing the size and morphology of the CCMO1 sample, revealing the Co/CoO core–shell particles embedded in a MgO matrix.

Magnetization measurements on the CCMO1 sample at 2 K in a magnetic field up to $\mu_0 H = \pm 5T$ are presented in Fig. 3. The hysteresis loop for zero-field cooling (ZFC, black line/symbols) of the CCMO1 sample is symmetric and thus without exchange bias. On the other hand, the hysteresis loop of the CCMO1 sample with field cooling (FC, red line/symbols) shown in Fig. 3 exhibits a clear asymmetry as compared with respect to ZFC data, yielding a giant exchange bias of about 2460 Oe (red square in Fig. 1). For comparison, we have also measured the ZFC and FC magnetization curves on a metallic sample $Co_{80}Mg_6O_{14}$ (called CCMO2), which is above the percolation threshold and has a large Co content (80%) or small Mg contents, and the obtained $H_E$ value (red squares) is shown in Fig. 1. Our results show that the formation of electronic percolation pathways between Co/CoO particles with increasing (decreasing) Co(Mg) content leads to a reduction of $H_E$.

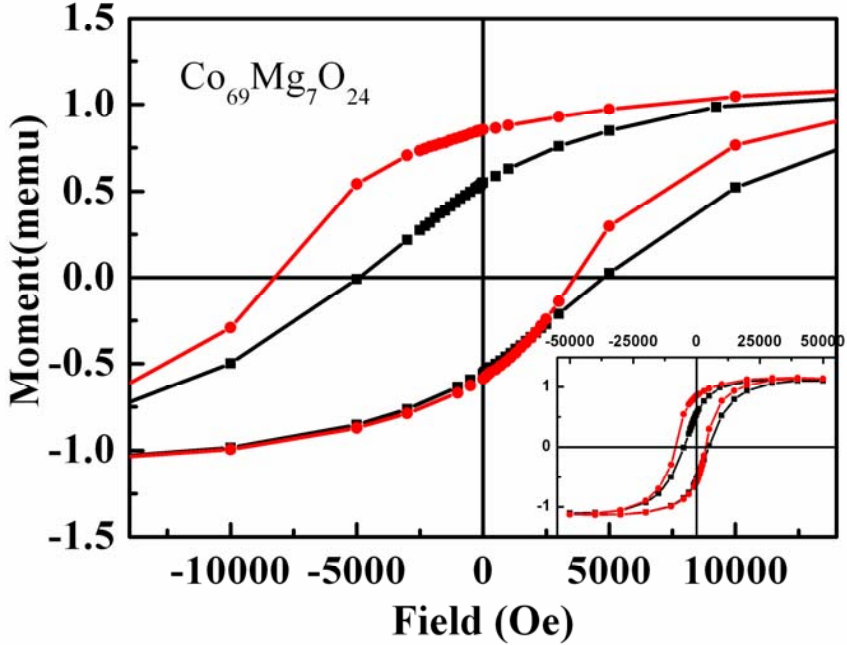

Fig. 3. The ZFC and FC magnetization curves of a CCMO1 sample at 2K. The red circles denote FC and the black squares denote ZFC data, respectively. The insert shows the whole hysteresis loop.

The obtained giant $H_E$ in CCMO1 sample can be understood within the framework of the physics of a diluted antiferromagnet at the interface. The lattice structures of the nonmagnetic MgO matrix and the antiferromagnetic CoO shell are very similar. Thus, Mg atoms are good substituents for Co atoms in the CoO shell. Diluting the Co atoms in the CoO shell by Mg causes the rising of uncompensated (UC) spins density at the AFM CoO surface in the same way as reported in the case of multilayered films [20,23]. In Table 1, we compare exchange bias field $H_E$ and lattice parameter mismatch for Co/CoO core-shell particles embedded in different matrix. A sample consisting of Co/CoO particles embedded in $Al_2O_3$ with the largest lattice mismatch (42.6%) shows no obvious exchange bias [20]. An exchange bias of about 6 Oe was observed in the system of Co/CoO nanoparticles (~6nm diameter) embedded in AFM $Cr_2O_3$ [24], in which the lattice mismatch is also very large (15.7%). By comparison with $Cr_2O_3$, the exchange bias of Co/CoO particles in a NiO or $SiO_2$ matrix is much larger [25]. Nevertheless, the $H_E$ in those systems is far lower than that in our Co/CoO-MgO system, in which MgO and CoO share the same NaCl crystallographic structure and their lattice parameters differ by only 1.1% [22]. As indicated in the Table 1, the $H_E$ has a decreasing tendency with increasing mismatch between the matrix and CoO crystal lattice parameters. Therefore, the MgO matrix plays an important role in the enhancement of the exchange bias in our Co/CoO-MgO system. Besides the dilution of Co by Mg as the main factor, some other factors may also influence the exchange bias field in the Co/CoO-MgO granular film as well. For example, the shape of those particles is irregular, consequently, the interface between the FM core and the AFM shell in those particles is rough, which also contributes to a high $H_E$ [18,19,26]. Moreover, the size of those particles in our granular films is smaller than 5 nm (as shown in Fig. 2) which can be considered as very small and it has to be noted in that context that $H_E$ is inversely proportional to the size of the particle [13].

TABLE I. Comparison of $H_E$ and lattice parameter mismatch of Co/CoO core-shell particles embedded in different matrix materials.

| Matrix | Mismatch | $H_E$(kOe) | Reference |
|---|---|---|---|
| $Al_2O_3$ | 42.6% | 0 | [27] |
| $Cr_2O_3$ | 15.7% | 0.006 | [28] |
| $SiO_2$ | 14.7% | 0.55 | [29] |
| NiO | 1.4% | 1.87 | [29] |
| MgO | 1.1% | 2.46 | this paper |

According to the uncompensated spin mechanism, a giant $H_E$ should be associated with a large number of uncompensated spin. In our Co/CoO-MgO system, the uncompensated spin should strongly depend on the concentration of Mg doping within the CoO shell. It is well-known that the soft x-ray absorption spectrum at the 3d transition metal $L_{2,3}$ edges is element and valence sensitive [27,28]. The x-ray absorption magnetic circular dichroism (XMCD) signal at the transition metal $L_{2,3}$ edges is an extremely sensitive probe for the local environment of transition metal atoms [27,28]. Thus, we use this technique to distinguish between the FM signal originated from the AFM CoO shell and the FM signal originated from Co metal cluster, and try to give a conclusive insight to the uncompensated (UC) spin model.

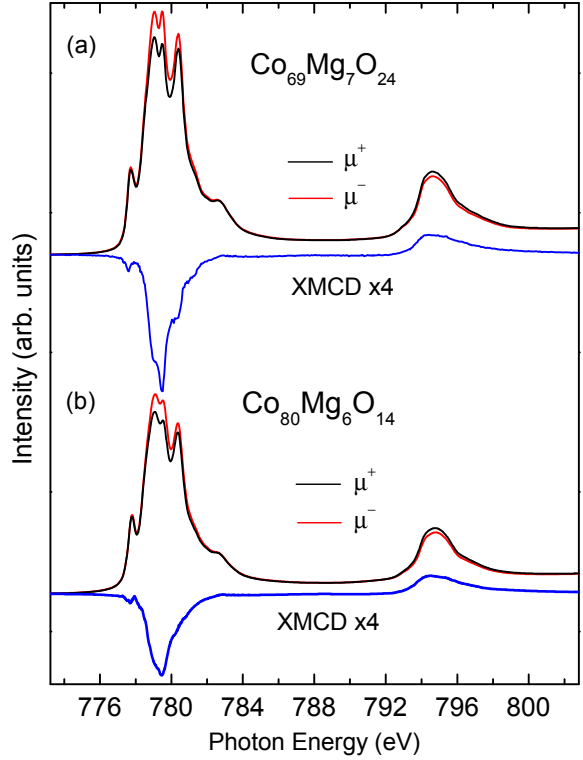

Fig. 4: Co-$L_{2,3}$ XMCD spectra of (a) sample CCMO1 and (b) sample CCMO2 measured at 80 K under a 5T magnetic field.

Fig. 4(a) depicts the Co-$L_{2,3}$ spectra of CCMO1 taken using circularly polarized light with the photon spin parallel ($\mu^+$ - black line) and antiparallel ($\mu^-$ - red line) with respect to the external magnetic field. The difference spectrum $\Delta\mu = \mu^+ - \mu^-$ ( i.e., the XMCD spectrum - blue line) is also shown in Fig. 4(a). The sharp multiplet structures which are observed in the $\mu^+$ and $\mu^-$ spectra are very similar to that of CoO [29], indicating the existence of divalent $Co^{2+}$ (i.e., CoO) in this material. Note that our important finding is that the XMCD spectrum in the CCMO1 sample has a multiplet spectral structure and its lineshape is very similar to that found in $LaCo_{0.5}Mn_{0.5}O_3$ [28], in which the Co ions have a divalent state and octahedral local crystal field symmetry.

This observation indicates unambiguously that the CoO shell contribute the XMCD signal and thus we do observe FM within the CoO shell.

The XMCD signal found from the CoO shell in the our Co/CoO-MgO system is quite surprising when considering the enormous research effort spent in the field of dilute FM in nonmagnetic semiconductors and insulators during the past decade, such as 3d TM ions doped into ZnO [30], $TiO_2$[31,32], and $BaTiO_3$[33]. A XMCD signal at the Co-$L_{2,3}$ edge of Co-$TiO_2$ was first observed by Kim et al.[34]. Later, it was realized that the Co-$L_{2,3}$ XMCD spectrum did not show the typical multiplet features originating from Co oxide, but rather from Co metal clusters exhibiting a structure-less spectral shape typical for Co metal. After failing to observe a XMCD signal at the TM-$L_{2,3}$ edge from oxidized TM, it was suggested that grain boundaries and related vacancies are at the origin of FM [30,35].

Now we turn our attention to the CCMO2 sample with a Co atomic ratio (80%) above the percolation threshhold. In Fig. 4(b), we present the absorption spectra $\mu^+$ and $\mu^-$ as well as the XMCD spectrum of CCMO2. Although the sharp multiplet spectral structures are still clearly visible in the $\mu^+$ and $\mu^-$ spectra, the multiplet structure in the XMCD spectrum nearly disappears. More precisely, one can see that the spectral line shape of the XMCD difference spectrum is nearly identical to that of Co metal as shown in Fig. 5(f) (magenta line). Here, the $\mu^+$ and $\mu^-$ XMCD spectra of Co metal in Fig. 5(d) were measured under the same experimental conditions and the overall spectral features are the same as measured using transmission geometry by Chen *et al.* [36]. For a Co content below the percolation threshold the samples become strongly charging and thus soft x-ray absorption spectra using the total electronic yield mode cannot be measured. This is another reason that we study samples at and above 69% Co content.

From the above observation, we can conclude that the sharp multiplet spectral features in the $\mu^+$ and $\mu^-$ absorption spectra of the CCMO2 sample are originating from the CoO shell, whereas the spectral structure in XMCD difference spectrum with a spectral shape typical for Co metal is essentially originated by the Co metal ions in the core clusters. This is exactly the same observation as in the case of Co clusters doped into $TiO_2$ [32]. The magnitude of XMCD difference signal is reduced from 7.2% to 4.6% when going from Fig. 4(a) to Fig. 4(b) as the CoO shell does not contribute to the XMCD signal in the CCMO2 sample. For such a large Co content, the size of $H_E$ are close to the classical values of compacted Co/CoO particles [37].

In order to estimate the amount of the CoO contribution to XMCD spectra for the CCMO1 sample, we have calculated the multiplet spectrum of the CoO shell using full multiplet calculations including crystal field interactions and covalence. Since the XMCD difference spectrum of Co metal does not exhibit any multiplet features, we make use of the experimental XMCD spectrum of Co metal measured under the same experimental conditions. Thus, we can simply simulate the experimental spectra of CCMO1 sample by a superposing the calculated spectra for CoO (Fig. 5c) and the experimental Co metal spectra ($\mu^+$ (black line) and $\mu^-$ (red line) in Fig. 5(d)). In our calculations we use similar parameters as used for CoO [38] ($U_{dd}$=6.5, $U_{cd}$=7.7 eV, $\Delta$=6.5, $pd\sigma$=-1.2, Slater integrals reduced by 75% from atomic Hartree-Fock values). The simulated $\mu^+$ and $\mu^-$ spectra from such a superposition presented in Fig. 5(b) nicely reproduce the experimental spectra given in Fig. 5(a).

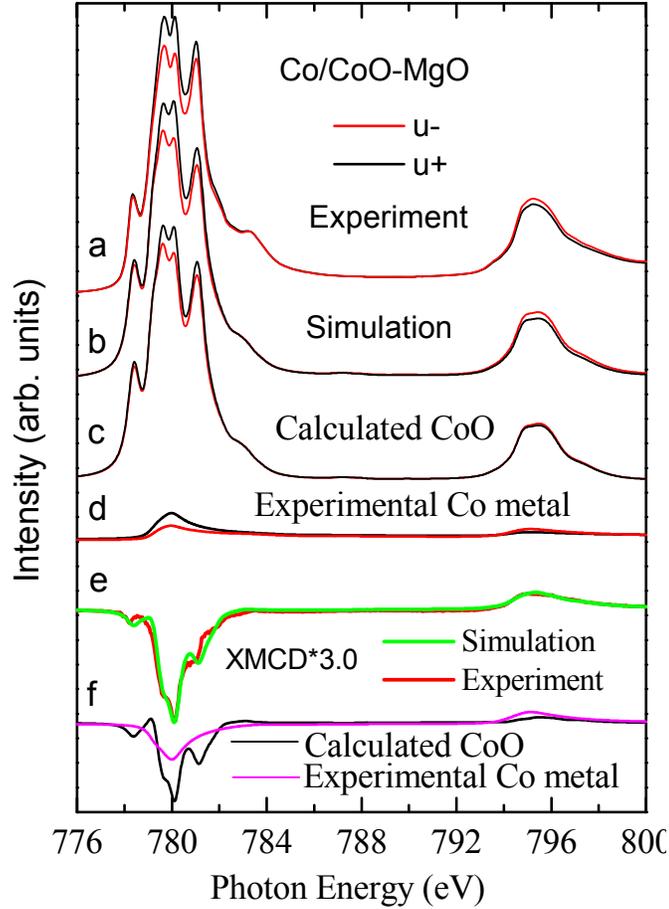

Fig. 5. (a) Experimental $\mu^+$ and $\mu^-$ spectra of CCMO1. (b): Simulated $\mu^+$ and $\mu^-$ spectra by a superposition of the calculated data for CoO shown in (c) and the experimental data for Co metal shown in (d). (e) Comparison of the experimental XMCD spectrum of CCMO1 and the simulated XMCD spectrum by a superposition of the calculated XMCD from CoO shell (black line shown in(f)) and experimental Co metal XMCD shown in (magenta line shown in (f)).

Fig. 5(e) also shows the nice agreement between the experimental (red line) and the simulated (green line) XMCD spectra. The latter is resulting from a superposition of the calculated XMCD spectrum for CoO shell (black line shown in (f)) and the experimental XMCD spectrum from Co metal (magenta line shown in (f)). From the above simulation we can estimate that about 70% of the XMCD spectrum is contributed by the CoO shell. However, this does not mean that CoO shell contribute really 70% ordered FM moment, since the effective electron escape depth in our total electron yield XAS spectrum is only in few nm. Our theoretical simulation further confirms that the FM signal from the CoO shell.

In summary, a series of Co/CoO-MgO nano-granular films were deposited on Si(100) substrates by using magnetron sputtering, exhibiting small isolated ferromagnetic Co clusters covered with a CoO shell embedded in a MgO matrix. A distinct percolation threshold with a sharp decrease of the resistance around 69% Co atomic ratio has been found, and a giant exchange bias field ($H_E \approx 2460$ Oe at T=2K) was obtained close to this critical percolation threshold. X-ray absorption magnetic circular dichroism at the Co-$L_{2,3}$ shows a clearly ferromagnetic signal originating from the nominally antiferromagnetic CoO shell for sample with Co content at the threshold. Decreasing the Mg-impurities will reduce the FM signal from CoO shell (namely the uncompensated spin density) and the size of $H_E$, therefore our work provides direct support for the uncompensated (UC) spin model.

**Acknowledgment**

The work was supported by the National Key Project for Basic Research of China (Grant No. 2010CB923404 and 2011CB922101) and PAPD. X.W. acknowledges support by NSFC under Grants No. 91122035, 11174124,11204124 and 10974082. XMCD experiments were performed at the BL29 Boreas beamline at the ALBA Synchrotron Light Facility with the collaboration of ALBA staff.

**References:**

[1] W. H. Meiklejohn and C. P. Bean, Phys. Rev. **102**, 1413 (1956).

[2] B. Dieny, V. S. Speriosu, S. Metin, S. Parkin, B. A. Gurney, P. Baumgart, and D. R. Wilhoit, J. Appl. Phys. **69**, 4774 (1991).

[3] B. Dieny, V. S. Speriosu, S. S. P. Parkin, B. A. Gurney, D. R. Wilhoit, and D. Mauri, Phys. Rev. B **43**, 1297 (1991).

[4] B. Dieny, J. Magn. Magn. Mater. **136**, 335 (1994).

[5] V. Skumryev, S. Stoyanov, Y. Zhang, G. Hadjipanayis, D. Givord, and J. Nogues, Nature **423**, 850 (2003).

[6] J. Nogués, J. Sort, V. Langlais, V. Skumryev, S. Suriñach, J. S. Muñoz, and M. D. Baró, Phys. Rep. **422**, 65 (2005).

[7] V. Laukhin, V. Skumryev, X. Martı´, D. Hrabovsky, F. Sa´nchez, M.V.

Garcı́a-Cuenca, C. Ferrater, M. Varela, U. Lüders, J. F. Bobo, and J. Fontcubertal , Phys. Rev. Lett. **97**, 227201 (2006).

[8] M. Ali, P. Adie, C. H. Marrows, D. Greig, B. J. Hickey, and R. L. Stamps, Nat. Mater. **6**, 70 (2007).

[9] H. Béa, M. Bibes, F. Ott, B. Dupé, X. H. Zhu, S. Petit, S. Fusil, C. Deranlot, K. Bouzehouane, and A. Barthélémy, Phys. Rev. Lett. **100**, 017204 (2008).

[10] J. de la Venta, M. Erekhinsky, S. Wang, K. G. West, R. Morales, and I. K. Schuller, Phys. Rev. B **85**, 134447 (2012).

[11] E. Lage, C. Kirchhof, V. Hrkac, L. Kienle, R. Jahns, R. Knöchel, E. Quandt, and D. Meyners, Nat. Mater. **11**, 523 (2012).

[12] S. Laureti, S. Y. Suck, H. Haas, E. Prestat, O. Bourgeois, and D. Givord, Phys. Rev. Lett. **108**, 077205 (2012).

[13] K. Takano, R. H. Kodama, A. E. Berkowitz, W. Cao, and G. Thomas, Phys. Rev. Lett. **79**, 1130 (1997).

[14] H. Ohldag, T.J. Regan, J. Stöhr, A. Scholl, F. Nolting, J. Lüning, C. Stamm, S. Anders, R.L. White, Phys. Rev. Lett. 87, 247201 (2001)

[15] H. Ohldag, A. Scholl, F. Nolting, E. Arenholz, S. Maat, A.T. Young, M. Carey, J. Stöhr, Phys. Rev. Lett. 91, 017203 (2003).

[16] Y. Ijiri, J. A. Borchers, R. W. Erwin, S. H. Lee, P. J. van der Zaag, and R. M. Wolf, Phys. Rev. Lett. **80**, 608 (1998).

[17] Jr. W.J. Antel, F. Perjeru, G.R. Harp, Phys. Rev. Lett. 83, 1439 (1999)

[18] N.C.Koon, Phys. Rev. Lett. **78**, 4865 (1997).

[19] C. Dufour, M. R. Fitzsimmons, J. A. Borchers, M. Laver, K. L. Krycka, K. Dumesnil, S. M. Watson, W. C. Chen, J. Won, and S. Singh, Phys. Rev. B **84**, 064420 (2011).

[20] Jung-Il Hong, Titus Leo, David J. Smith, and Ami E. Berkowitz, Phys. Rev. Lett. **96**, 117204 (2006).

[21] J. Keller, P. Milte´nyi, B. Beschoten, G. Güntherodt, U. Nowak, and K. D. Usadel, Phys. Rev. B **66**, 014431 (2002).

[22] Kentaro Takano, R. H. Kodama, and A. E. Berkowitz, Phys. Rev. Lett. **79**, 6 (1997).

[23] L. Titus, H. Jung-Il, B. E. Ami, and J. E. David, J. Appl. Phys. **102**, 123904 (2007).

[24] D. Tobia, E. Winkler, R. D. Zysler, M. Granada, H. E. Troiani, and D. Fiorani, J. Appl. Phys, **106**, 103920 (2009).

[25] P. Céline, M. Robert, B. Ariel, and N. Lucien, J. Appl. Phys **100**, 033907 (2006).

[26] D. Lederman, J. Nogue´s, and Ivan K. Schuller，Phys. Rev. B **56**, 2332 (1997).

[27] J. A. Moyer, C. A. F. Vaz, D. A. Arena, D. Kumah, E. Negusse, and V. E. Henrich, Phys. Rev. B **84**, 054447 (2011)

[28] T. Burnus, Z. Hu, H. H. Hsieh, V. L. J. Joly, P. A. Joy, M. W. Haverkort, Hua Wu, A. Tanaka, H.-J. Lin, C. T. Chen, and L. H. Tjeng, Phys. Rev. B **77**, 125124 (2008)

[29] S. I. Csiszar, M.W. Haverkort, Z. Hu, A. Tanaka, H. H. Hsieh, H.-J. Lin, C. T. Chen, T. Hibma, and L. H. Tjeng, Phys. Rev. Lett. 95, 187205 (2005)

[30] T. Thomas, G. Milan, S. Gisela, J. Gerhard, B. Sebastian, and G. Eberhard, New Journal of Physics 10, 055009 (2008)

[31] M.Yuji,M. Makoto, S. Tomoji, H. Tetsuya, F. Tomoteru, K. Masashi, A. Parhat, C.Toyohiro,K. Shin-ya, and K. Hideomi, Science 291,854 (2001),

[32] T. Ohtsuki, A. Chainani, R. Eguchi, M. Matsunami, Y. Takata, M. Taguchi, Y. Nishino, K. Tamasaku, M. Yabashi, T. Ishikawa, M. Oura, Y. Senba, H. Ohashi, and S. Shin, Phys. Rev. Lett. 106,047602(2011)

[33] Tanushree Chakraborty, Sugata Ray, Mitsuru Itoh, Phys. Rev. B 83, 144407 (2011)

[34] J.Y. Kim, J.H. Park, B.G. Park, H.J. Noh, S.J. Oh, J. S.Yang, D.H. Kim, S. D. Bu, T.W. Noh, H.J. Lin, H.H. Hsieh, and C.T. Chen, Phys. Rev. Lett. 90, 017401 (2003)

[35] B. B. Straumal, A. A. Mazilkin, S. G. Protasova, A. A. Myatiev, P. B. Straumal, G. Schütz, P. A. van Aken, E. Goering, and B. Baretzky, Phys. Rev. B 79, 205206 (2009)

[36] C.T. Chen et al. Phys. Rev. Lett. **75**, 152 July (1995)

[37] D. L. Peng, K. Sumiyama, T. Hihara, S. Yamamuro, and T. J. Konno, Phys. Rev. B **61**, 3103–3109 (2000)

[38] G. Ghiringhelli, L.H. Tjeng, A. Tanaka, O. Tjernberg, T. Mizokawa, J.L. de Boer,

N.B. Brookes, Phys. Rev. B 66, 075101 (2002).